\documentclass[a4paper,fleqn,usenatbib]{mnras}

\usepackage[T1]{fontenc}
\usepackage{ae,aecompl}

\usepackage{graphicx}	% Including figure files
\usepackage{amsmath}	% Advanced maths commands
\usepackage{amssymb}	% Extra maths symbols
\usepackage{array}

\newcommand{\erg}{erg cm$^{-2}$ s$^{-1}$} % unit of Flux
\newcommand{\xm}{\emph{XMM-Newton}}
\newcommand{\sw}{\emph{Swift}/XRT}
\newcommand{\nus}{\emph{NuSTAR}}
\newcommand{\sax}{SAX J1748.9--2021}

\title[Reflection Spectra of \sax]{Study of the reflection spectra of \sax}

\author[R. Sharma et al.]{Rahul Sharma$^{1}$\thanks{E-mail: rahul1607kumar@gmail.com},
Chetana Jain$^{2}$
and Anjan Dutta$^{1}$
\\
$^{1}$Department of Physics and Astrophysics, University of Delhi, Delhi 110007, India\\
$^{2}$Hansraj College, University of Delhi, Delhi 110007, India\\
}

\date{Accepted XXX. Received YYY; in original form ZZZ}

\pubyear{2018}

\begin{document}
\label{firstpage}
\pagerange{\pageref{firstpage}--\pageref{lastpage}}
\maketitle

\begin{abstract}
We report the spectral analysis of accretion powered millisecond X-ray pulsar \sax\ from the \nus\ observation made during its 2015 outburst. The spectra displayed a broad emission line at $\sim 6.5$ keV with line width of $\sim 0.5$ keV and excess above $\sim 20$ keV due to Compton hump. The persistent emission of \sax\ is described by a combination of soft thermal component with $kT =0.62^{+0.03}_{-0.05}$ keV and thermally Comptonized component with $kT_e=2.50^{+0.06}_{-0.03}$ keV reflected from the disc with reflection fraction of $0.30^{+0.08}_{-0.04}$. 
We have modeled the reflection spectrum with self-consistent model \texttt{relxillCP} and have found the inclination of the accretion disc to be $32.3^{^\circ +4.8}_{-4.7}$ and log $\xi$ is equal to $3.05^{+0.40}_{-0.34}$. We have obtained an upper limit of 57 km for inner disc radius; and the estimated magnetic field strength at the poles is less than $3.4 \times 10^9$ G. This kind of detailed investigation of reflection spectrum of \sax, especially above 10 keV, has been achieved only because of high statistics \nus\ data of the source.  
\end{abstract}


%%%%%%%%%%%%%%%%%%%%%%%%%%%%%%%%%%%%%%%%%%%%%%%%%%

%%%%%%%%%%%%%%%%% BODY OF PAPER %%%%%%%%%%%%%%%%%%

\section{Introduction}

Accretion powered millisecond X-ray pulsars (AMXPs) are rapidly rotating neutron stars (NS) which are known to accrete from a low mass companion \citep[see e.g.,][for reviews]{Poutanen2006, Patruno2012}. The magnetic field of AMXPs generally lie in the range $\sim 10^7-10^9$ G \citep{Mukherjee2015}. The magnetic field of AMXPs is strong enough to truncate the accretion disc far from the stellar surface and channel the accreting material towards the magnetic polar caps of the NS \citep[e.g.,][]{Cackett2009, Papitto2009, Papitto2013, Pintore}. The X-rays emitted by the release of gravitational potential energy above the NS surface is then observed in the form of pulsations \citep[see e.g.,][]{Ghosh}.

\sax\ is a transient AMXP which was discovered in 1998 with \emph{Beppo-SAX} \citep{intZand1999}. It is located in the globular cluster NGC 6440 at a distance of $\sim 8.5$ kpc \citep{Ortolani1994, Kuulkers2003}. This source shows intermittent pulsations at $\sim 442$ Hz and it has an orbital period of $\sim 8.76$ h \citep{Altamirano2008, Patruno2009, Sanna}. The mass and radius of the companion star have been estimated to lie in the range $0.70-0.83 M_{\sun}$ and $0.86-0.90 R_{\sun}$, respectively \citep{Cadelano}. 
\sax\ also went into the outbursts in 2001, 2005, 2010, 2015 and 2017 \citep{intZand2001, Markwardt2005, Patruno2010,  Kuulkers-detect, Negoro2017}. During the outburst of 2015, source was in the soft state and its broadband spectrum consisted of two soft thermal components, a cold thermal Comptonization ($\sim 2$ keV) and an additional hard X-ray emission modeled using power-law \citep{Pintore}. During the outburst of 2017, source was observed in low hard state, with average spectral properties consistent with a soft thermal component plus a hot thermal Comptonization \citep{Pintore2018}.

The reflection spectrum from accretion disc has been studied in AMXPs such as SAX J1808.4-3658 \citep[e.g.,][]{Cackett2010, Wilkinson2011}, HETE J1900.1-2455 \citep{Cackett2010, Papitto2013}, IGR J17511-3057 \citep{Papitto2010} and Aql X-1 \citep{Ludlam2017c}. Generally, the reflection process results in a broad Fe K$\alpha$ emission line around 6.4--7.0 keV and a broad Compton back-scattering hump between $\sim 10-30$ keV \citep{Ross1999}. Modeling of these features can provide information about the disc geometry, like the inner disc radius and inclination of disc \citep{Fabian1989, Fabian2010}. 

Previous studies of \sax\ lack a proper investigation of reflection spectrum \citep{intZand1999, Pintore, Pintore2018}.
During the outburst of 2015, \citet{Pintore} modeled the broad emission feature using \texttt{diskline} \citep{Fabian1989} and they also reported that the reflection models, \texttt{reflionx} \citep{Ross2005} and \texttt{rfxconv} \citep{Kolehmainen2011}, were not able to give a stable fit because of low statistic data above 10 keV and hence constrained the parameters.
In this work, we present the broadband reflection spectroscopy of the AMXP source \sax\ by making use of the \nus\ observation made during the outburst of 2015.

\section{Observations and data analysis}

The Nuclear Spectroscopic Telescope ARray \citep[\nus;][]{Harrison} has been used to study the reflection features of many LMXBs \citep[e.g.,][]{Mondal2017, Ludlam2017b, Ludlam2017a, Wang2017}. 
The \nus\ 
mission consists of two telescopes, which focus X-rays between 3 and 79 keV onto two identical focal planes (FPMA and FPMB). It has a field of view (FOV) of $12' \times 12'$ and an angular resolution of $18''$ FWHM. 
Because of the broad energy coverage, it can simultaneously observe the broad emission line and the Compton hump.

\begin{figure}
\centering
\includegraphics[width=0.9\columnwidth]{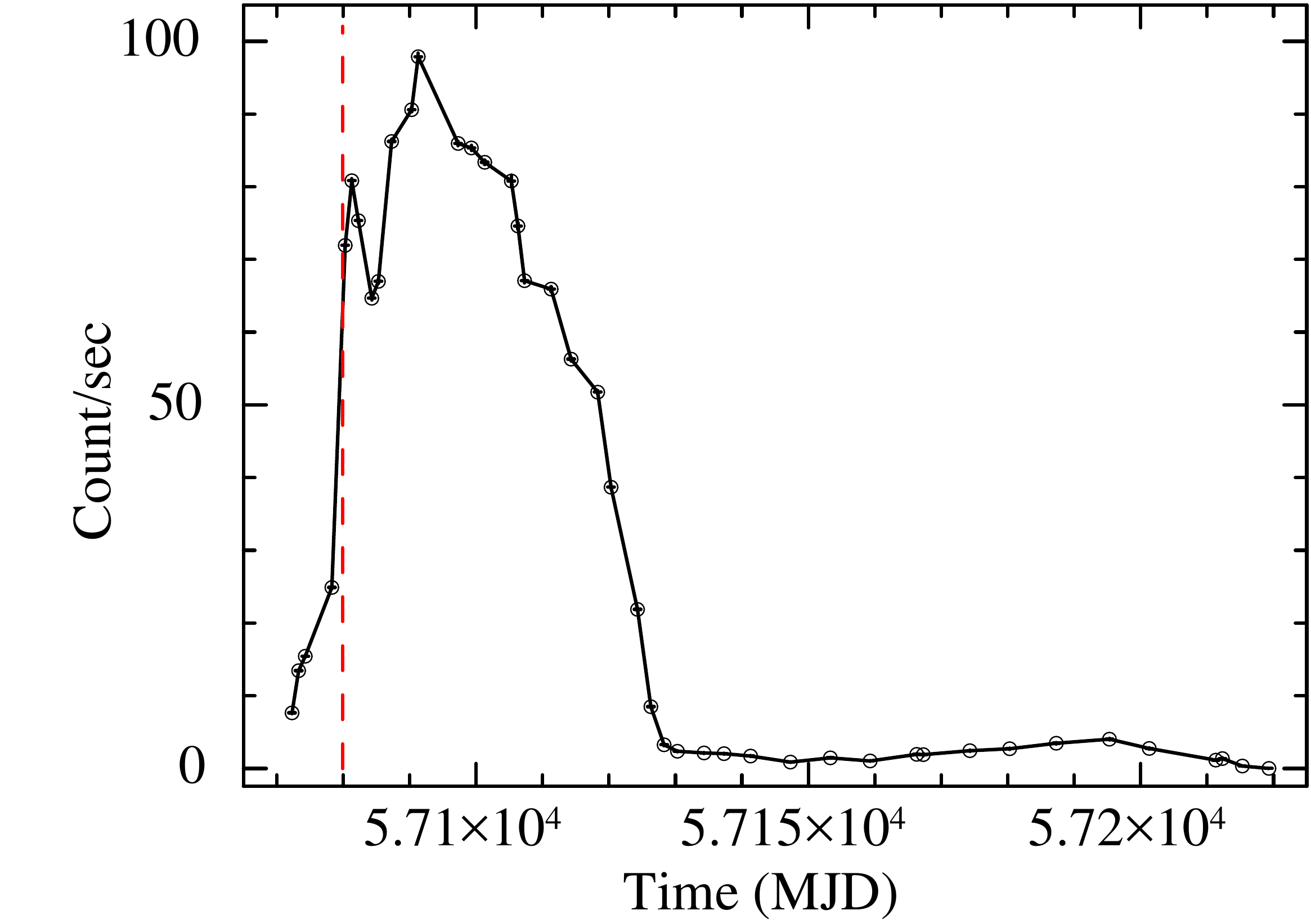}
 \caption{0.5--10 keV \sw\ light curve of the 2015 outburst of \sax. The red dashed line indicates the epoch of the \nus\ observation.}
\label{xrt-lc}
\end{figure}

\begin{figure}
\centering
\includegraphics[width=0.9\columnwidth]{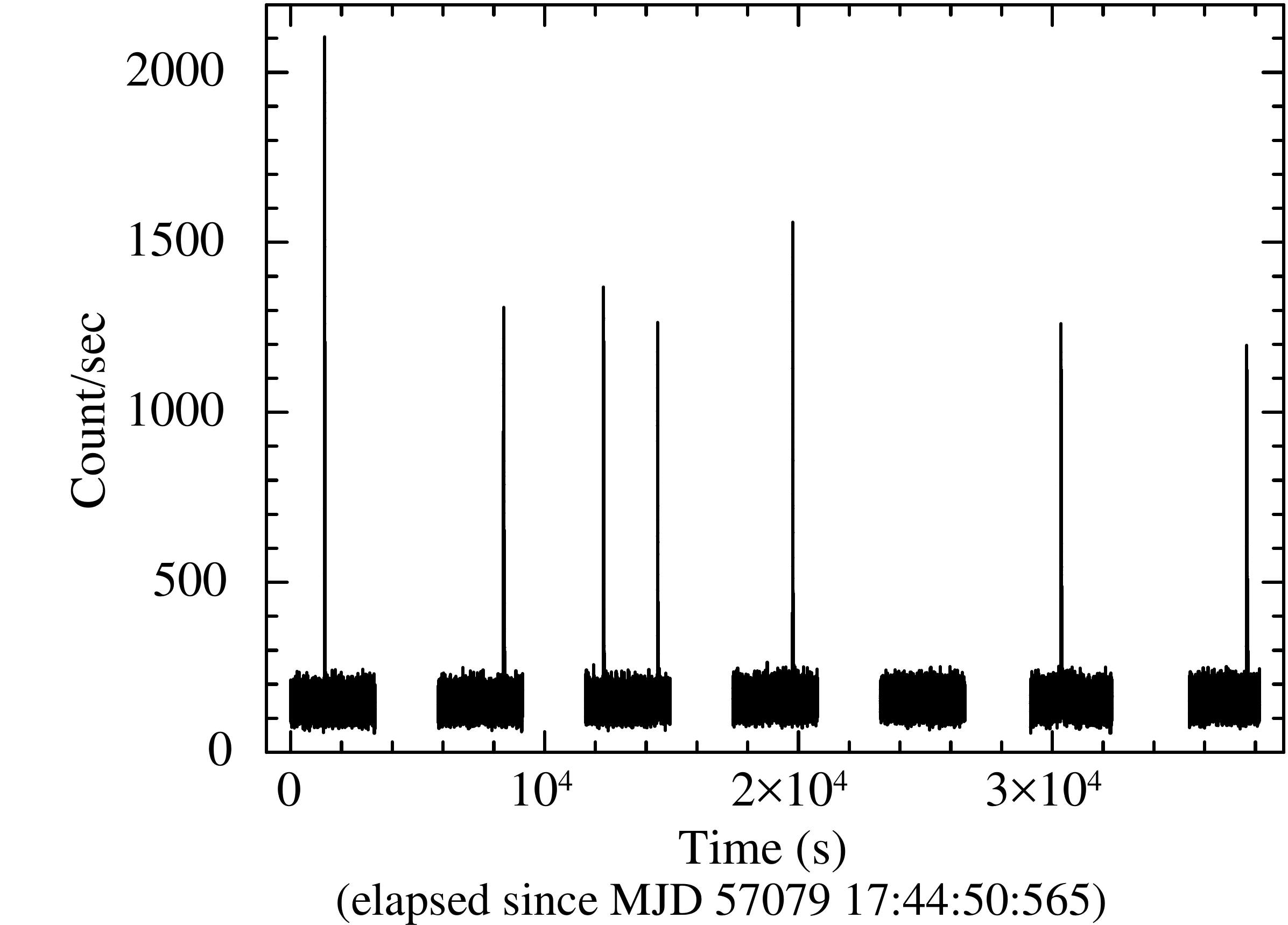}
 \caption{3--78 keV lightcurve of \sax\ extracted from the \nus\ observation.}
\label{lc}
\end{figure}

\sax\ was observed with \nus\ on 2015 February 26 for a total exposure time of 18 ks (obs-id : 90001002002), during the initial phase of the 2015 outburst. 
Figure \ref{xrt-lc} shows the total outburst lightcurve of the source as observed by \sw\ during 2015. The epoch of \nus\ observation has been marked with a red vertical dash line.
We have used the most recent \nus\ analysis software \textsc{nustardas v.1.8.0} distributed with  \textsc{heasoft} version 6.22 and the latest calibration files (version  20170817) for reduction and analysis of the \nus\ data. We used the task \textsc{nupipeline} to generate calibrated and screened event files. A circular region of radius $120''$ centered at the source position was used to extract the source events. Background events were extracted from a circular region of same size away from the source. The task \textsc{nuproduct} was used to generate the lightcurves, spectra and response files. The FPMA/FPMB light curves were background-corrected and summed using \textsc{lcmath} and is shown in Figure \ref{lc}. In this observation, seven type-I thermonuclear X-ray bursts are observed. 
The persistent emission spectra were extracted by excluding the bursts (removed data between 10 sec before the burst start and 200 sec after the burst start). The spectra were grouped to give a minimum of 25 counts/bin.  

%-------------------------------------------------------------------

\section{Spectral Analysis}

We used the \textsc{xspec} version 12.9.1m \citep{Arnaud} for the spectral fitting. To model  interstellar absorption, \texttt{tbabs} was used, with the abundances set to \texttt{wilm} \citep{Wilms} and cross sections set to \texttt{vern} \citep{Verner}. We have fitted the persistent spectra extracted from \nus-FPMA and \nus-FPMB, simultaneously. For each spectrum, we added a multiplicative constant to account for the different cross-calibrations of the instruments and fixed this constant to 1 for \nus-FPMA. \sax\ was significantly detected upto 40 keV above background. Therefore, we have fitted the spectra in the energy range 3--40 keV. All spectral uncertainties and upper-limits are given at 90\% confidence unless specified.

We fitted the spectrum with absorbed thermally Comptonized component \texttt{nthcomp} \citep{Zdziarski, Zycki}. 
The \texttt{nthcomp} model pictures that the hot electrons Compton upscatter the seed photons with a (quasi)blackbody spectrum e.g., from a NS surface/boundary layer or disc blackbody spectrum. 
As \nus\ data extends down to 3 keV only, the value of $N_H$ cannot be constrained from our fit. Therefore, we fixed $N_H$ to $0.58 \times 10^{22}$ cm$^{-2}$ \citep{Pintore}.
The \texttt{nthcomp} model provided unacceptable fit ($\chi^2_{red}=1.42$ for 1055 dof) and residuals showed the excess in tail above 20 keV and broad emission feature in 6--7 keV (Figure \ref{nthcomp}). The broad emission line and excess above $\sim 30$ keV was also observed with \xm$+$\emph{INTEGRAL} data during this outburst \citep{Pintore}. 

\begin{figure}
\centering
\includegraphics[width=0.9\columnwidth]{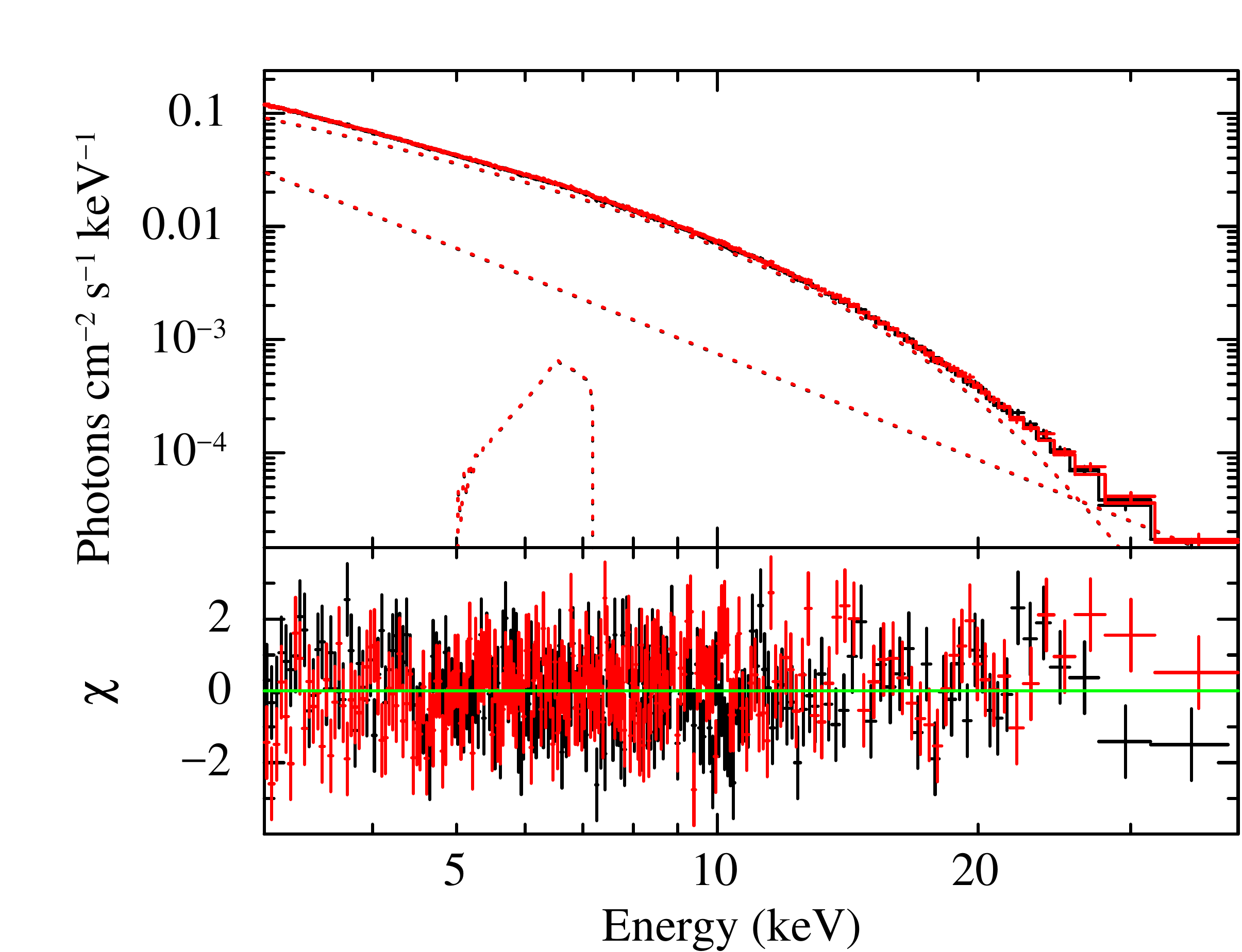}
 \caption{The persistent spectrum of \sax\ fitted with \texttt{tbabs$\times$(nthcomp+powerlaw+diskline)} model. The figure has been rebinned for representation purpose.}
\label{modelA}
\end{figure}

We initially modeled the 6--7 keV broad feature with Gaussian emission line, found at $6.52^{+0.14}_{-0.16}$ keV (line width $\sim 0.5$ keV) with equivalent width (EW) of $31^{+20}_{-9}$ eV and excess above 20 keV with \texttt{powerlaw} (Table \ref{table1}).  
The powerlaw component obtained was steep with photon index $\Gamma = 3.1 \pm 0.2$. We then replaced the Gaussian component with  a relativistically smeared reflection profile, \texttt{diskline}. The introduction of a \texttt{diskline} model did not provide significant improvement over \texttt{Gaussian}, ($\Delta \chi^2=3$, see Table \ref{table1}). Moreover, the fit with \textsc{diskline} model was insensitive to the parameter like inclination angle, outer radii and emissivity. Therefore, we fixed them to $44^{\circ}$, $10^5~R_g$ ($R_g = GM/c^2$ is the gravitational radii, which is $\simeq 2.07$ km for $M=1.4 M_{\sun}$) and --2.7, respectively \citep{Pintore}. The Fe K emission feature is observed at $6.50^{+0.12}_{-0.15}$ keV with EW of $\sim 32^{+27}_{-10}$ eV, produced in the disc at a distance of $R_{in} < 34 ~R_g$ (Table \ref{table1} and Figure \ref{modelA}). 

\subsection{Reflection Spectrum}

\begin{figure}
\centering
\includegraphics[width=0.9\columnwidth]{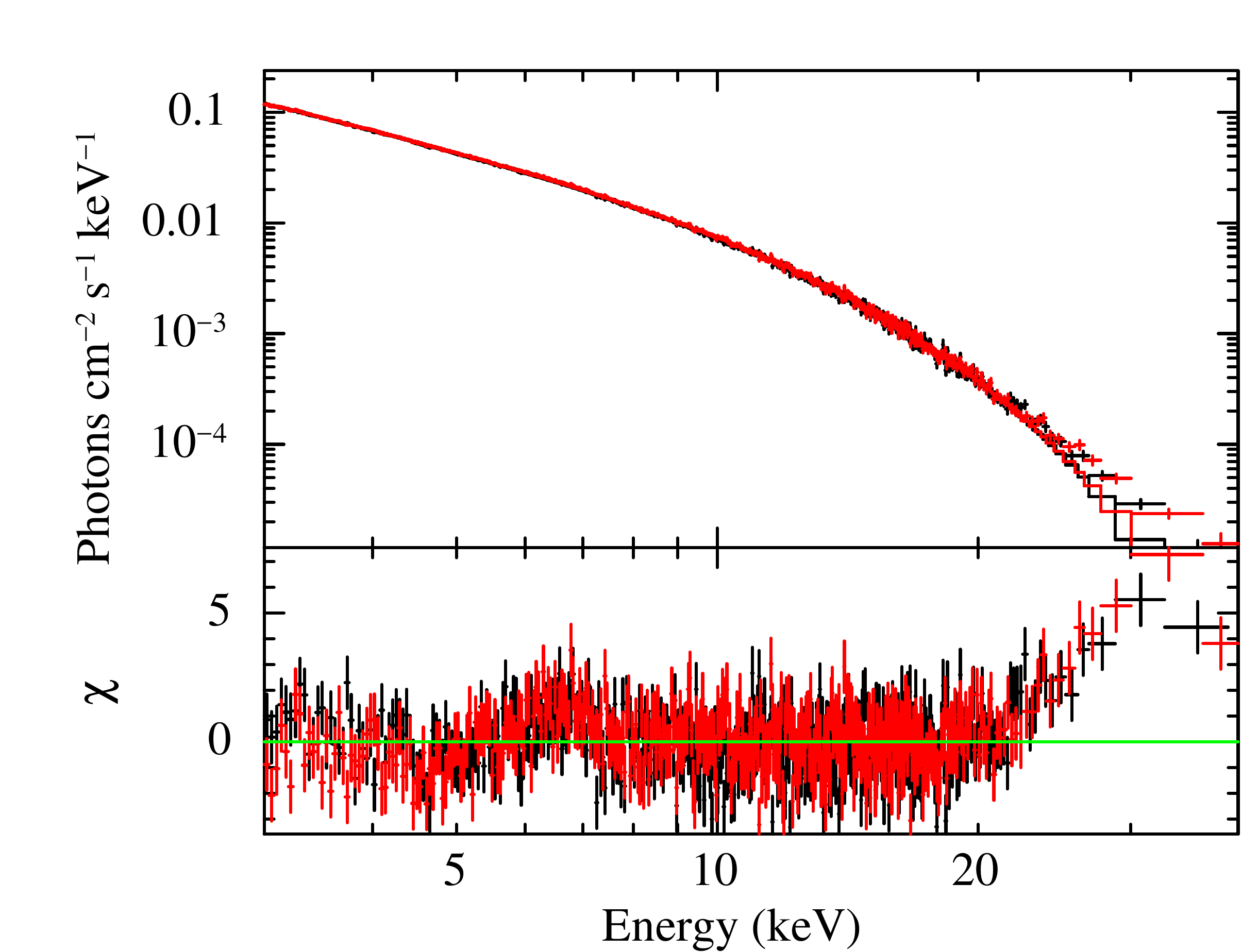}
 \caption{ \nus\ spectrum modeled with \texttt{nthcomp} Comptonized component. The figure has been rebined for representation purpose. The residuals shows the broad feature at 6--7 keV and large excess above 20 keV. }
\label{nthcomp}
\end{figure}

The high energy tail above 20 keV can be due to Compton reflection hump. The presence of the broad emission line and strong Compton reflection hump (see Fig. \ref{nthcomp}) suggests for the fitting of broad-band self-consistent reflection model. 
Therefore, we applied the self-consistent reflection model \texttt{relxill} \citep{Garcia2014, Dauser2014} v. 1.0.2 that calculates disc reflection features due to irradiation of the accretion disc by a broken power-law emissivity. It combines the reflection code \texttt{xillver} \citep{Garcia2010, Garcia2013} and the relativistic ray tracing code \texttt{relline} \citep{Dauser2010, Dauser2013}, in which reflection spectrum is chosen for each relativistically calculated emission angle. This model features higher spectral resolution and updated atomic data compared to other models. We used the \texttt{relxillCP} \citep{Garcia2018} model which describes the reflection from the physical Comptonization continuum calculated using \texttt{nthcomp}.
We used the single emissivity profile ($\propto r^{-q}$) and fixed $q=3$ \citep{Wilkins2012, Cackett2010}. We also fixed the outer radius $R_{out} = 1000 R_g$, as the sensitivity of the reflection fit decreases with increasing outer disc radius. 
The dimensionless spin parameter $a$ can be calculated from the spin frequency using the relation $a = 0.47/P[ms]$ \citep{Braje2000}. Using the spin frequency $\nu =442$ Hz, we fixed $a$ at 0.208. 

Initially, we added the \texttt{relxillCP} to \texttt{nthcomp} by fixing the $refl_{frac}$ to negative value to get only reflected spectrum, such that, the \texttt{nthcomp} represents the direct coronal emission component and \texttt{relxillCP} the reflected component. We tied the power-law photon index, $\Gamma$, and electron temperature, $kT_e$, of the \texttt{relxillCP} to that of the respective \texttt{nthcomp} parameters. We also found that the addition of a soft component, \texttt{bbodyrad} or \texttt{diskbb} \citep{Mitsuda}, improved the fit significantly (with F-test probability of $\sim 10^{-7}$), giving the resultant $\chi^2_{red}=1.083$ for 1049 dof. 

\begin{figure}
\centering
\includegraphics[width=0.9\columnwidth]{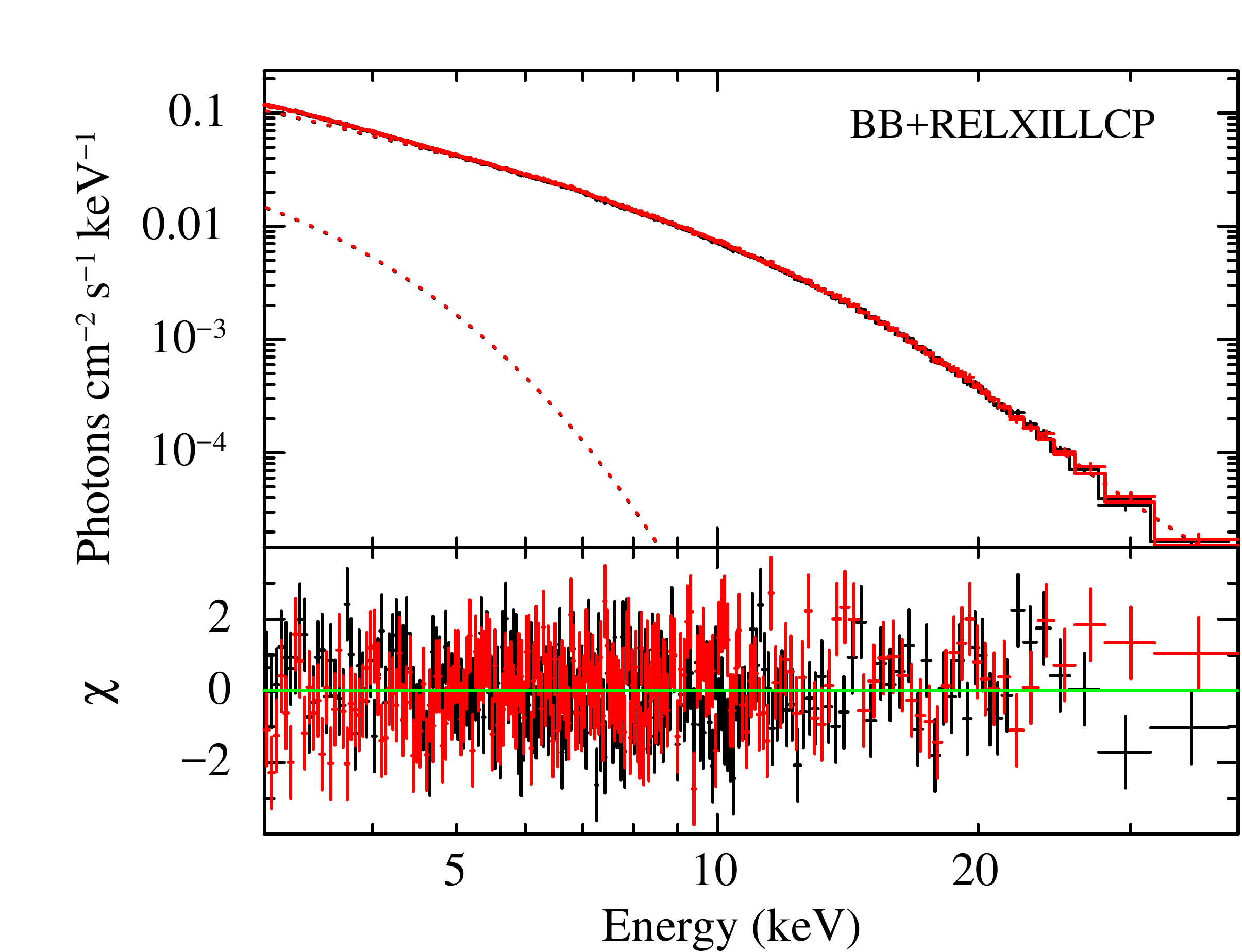}
 \caption{The persistent spectrum of \sax\ fitted with self consistent reflection model, \texttt{relxillCP}. The figure has been rebinned for representation purpose.}
\label{modelB}
\end{figure}

Next, we replaced the \texttt{nthcomp} with \texttt{relxillCP}, since it describes both the illuminating and reflection emission and kept $refl_{frac}>0$. The $refl_{frac}$ is defined as the ratio of the intensity illuminating the disc to the intensity reaching the observer or infinity \citep{Dauser2016}. A soft thermal component improved the fit significantly. We found that the \texttt{relxillCP} parameters were independent of the chosen thermal component, \texttt{diskbb} or \texttt{bbodyrad} and both components gave similar fit statistics and hence we will discuss the results only with \texttt{bbodyrad}. 
The best fit parameters obtained are given in Table \ref{para} and the resultant spectrum is shown in Figure \ref{modelB}. We obtained the reflection fraction of $0.30^{+0.08}_{-0.04}$, iron abundance of disc atmosphere in Solar $A_{Fe} = 2.6^{+2.3}_{-1.6}$, disc inclination angle $i=32.3^{\circ +4.8}_{-4.7}$ and upper limit on inner radius of accretion disc $R_{in} < 5.2~R_{ISCO}$, where $R_{ISCO}$ is the radius of inner-most stable circular orbit.

\begin{figure}
\centering
\includegraphics[width=0.9\columnwidth]{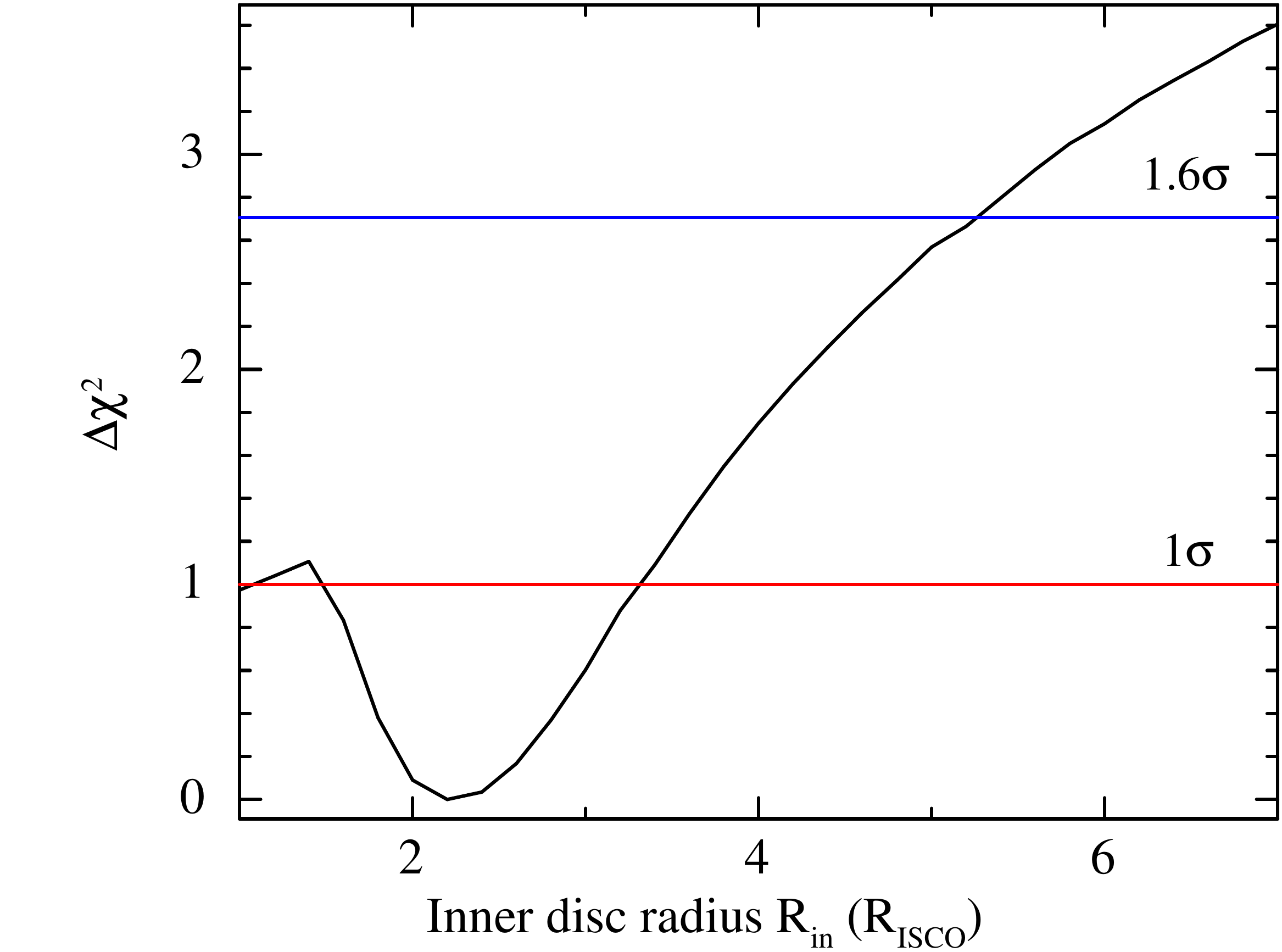}
 \caption{ The $\Delta \chi^2$ contour of the inner disc radius. The red and blue horizontal lines represent the $1\sigma$ and $1.6\sigma$ significance level, respectively.}
\label{con-rin}
\end{figure}

We computed $\Delta \chi^2$ for inner disc radius using steppar in \textsc{xspec}. The resultant $\Delta \chi^2$ while varying inner disc radius between $1~R_{ISCO}$ and $7~R_{ISCO}$ in steps of $0.2~R_{ISCO}$ is shown in the Figure \ref{con-rin}. In this figure, $1 \sigma$ and $1.6 \sigma$ significance levels are shown by horizontal lines. The inner disc radius can be constrained at only $1\sigma$ level with $R_{in}= 2.2^{+1.1}_{-0.7}~R_{ISCO}$.

%%%%%%%%%%%%%%%%%%%%%%%%%%%%%%%%%%%%%%%%%%%%%%%%%%%%%%%%%%%%

%%%%%%%%%%%%%%%%%%%%%%%%%%%%%%%%%%%%%%%%%%%%%%%%%%%%%%%%%%%%%

\begin{table}
\caption{Spectral parameters of \sax\ with phenomenological model and emission line modeled using \texttt{gaussian/diskline}.}
\centering
\resizebox{0.9\columnwidth}{!}{
\begin{tabular}{c c c c}
\hline \hline
&Parameters & \textsc{gaussian} & \textsc{diskline} \\
\hline
		& $\Gamma$	  & $1.90^{+0.04}_{-0.06} $ & $1.89^{+0.03}_{-0.04} $ \\[0.1ex]
		& $kT_{e}$ (keV)  & $2.58^{+0.04}_{-0.03} $ & $2.57 \pm 0.03 $ \\[0.1ex]
		& $kT_{seed}$ (keV)& $0.42^{+0.23}_{-0.18}$ & $0.42^{+0.21}_{-0.04}$ \\[0.1ex]
		& input\_type     & $0$ & $0$ \\[0.1ex]
		& Norm$_{NTHCOMP}$& $0.31^{+0.36}_{-0.18} $ & $0.30^{+0.23}_{-0.17} $ \\[0.1ex]

\hline
		& $\Gamma$ & $3.1 \pm 0.2 $ & $3.11 \pm 0.15 $ \\[0.1ex]
		& Norm$_{PL}$	  & $0.95^{+1.19}_{-0.51} $ & $0.97 \pm 0.41 $ \\[0.1ex]
\hline

		& $E_{line}$ (keV) & $ 6.52^{+0.14}_{-0.16} $ & $ 6.50^{+0.12}_{-0.15} $ \\[0.1ex]
		& $\sigma$ (keV)  & $ 0.48^{+0.18}_{-0.15} $ & - \\[0.1ex]
		& $betor10$	   & - & $-2.7^{fixed} $ \\[0.1ex]
		& $R_{in}$ ($R_{g}$)& - & $<34$ \\[0.1ex]
		& $R_{out}$ ($R_{g}$)& - & $10^{5 (fixed)}$ \\[0.1ex]
		& Inclination (deg) & - & $ 44^{fixed} $ \\[0.1ex]
		& EW (eV)	  & $31^{+20}_{-9}$ & $32^{+27}_{-10}$ \\[0.1ex]
		& Norm$_{line}$ (10$^{-4}$)& $ 7.2^{+2.7}_{-2.1} $ & $ 7.2^{+2.6}_{-1.8} $ \\[0.1ex]

\hline
		& $\chi^2_{red}$ (dof) & 1.093 (1050) & 1.090 (1050) \\
\hline
\end{tabular}}
\label{table1}
\end{table}

%%%%%%%%%%%%%%%%%%%%%%%%%%%%%%%%%%%%%%%%%%%%%%%%%%%%%%%%%%%%%%

\begin{table}
\caption{Spectral parameters of \sax\ with best fit model \texttt{tbabs$\times$(bbodyrad+relxillCP)}.}
\centering
\resizebox{1.08\columnwidth}{!}{
\begin{tabular}{c c l}
\hline \hline
Parameters & Value & Parameter details  \\\hline
$kT_{BB}$ (keV) & $0.62^{+0.03}_{-0.05} $ & The blackbody temperature\\[0.1ex]
 Norm	  & $217^{+81}_{-48} $ & Normalization of Blackbody\\[0.1ex]
\hline
$q$	 	  & $3^{fixed} $ & Emissivity index\\[0.1ex]
$a$		  & $0.208^{fixed}$ & Spin parameter\\[0.1ex]
$R_{in}$& $<5.2$ & Inner disc radius in $R_{ISCO}$\\[0.1ex]
& $<27.7$ & Inner disc radius in $R_{g}$\\[0.1ex]
$R_{out}$ ($R_{g}$)& $1000^{fixed}$ & Outer disc radius\\[0.1ex]
Inclination (deg) & $ 32.3^{+4.8}_{-4.7} $ & Disc inclination angle\\[0.1ex]
$A_{Fe}$	  & $ 2.6^{+2.3}_{-1.6} $ & Iron abundance in Solar\\[0.1ex]
log $\xi$	  & $ 3.05^{+0.40}_{-0.34} $ & Ionization parameter \\[0.1ex]
$\Gamma$	  & $1.82^{+0.07}_{-0.06} $ & Photon Index\\[0.1ex]
$kT_{e}$ (keV)  & $2.50^{+0.06}_{-0.03} $ & Comptonizing electrons temperature\\[0.1ex]
$refl_{frac}$	  & $0.30^{+0.08}_{-0.04}$ & Reflection fraction\\[0.1ex]
Norm$_{refl}$ (10$^{-3}$)& $ 6.98^{+1.3}_{-0.8} $ & Normalization of \texttt{relxillCP}\\[0.1ex]

\hline
$^\dagger$Flux$_{BB}$      & $0.23 \pm 0.05$ & Flux of blackbody component\\[0.1ex]
$^\dagger$Flux$_{relxill}$    & $3.38 \pm 0.08$ & Flux of relxillCP component\\[0.1ex]
$^\dagger$Flux$_{total}$   & $3.61 \pm 0.03$ & Total flux \\[0.1ex]
$^{\dagger\dagger}L_X$   & 3.1 & X-ray Luminosity \\[0.1ex]
\hline
$\chi^2_{red}$ (dof) & 1.083 (1049) & Fit statistics\\
\hline
\multicolumn{3}{l}{$\dagger$1.6--100 keV unabsorbed flux in unit of $10^{-9}$ \erg.}\\
\multicolumn{3}{l}{$^{\dagger\dagger}$1.6--100 keV unabsorbed luminosity in units of $10^{37}$ erg s$^{-1}$}\\
\multicolumn{3}{l}{(assuming distance of 8.5 kpc).}\\
\end{tabular}}
\label{para}
\end{table}

%-----------------------------------

%%%%%%%%%%%%%%%%%%%%%%%%%%%%%%%%%%%%%%%%%%%%%%%%%%%%%%%%%%%%%%

\section{Discussion and Conclusions}

In this work, we have performed the spectral analysis of AMXP \sax\ during the rise of its 2015 outburst with \nus\ observation. The persistent spectra can be well described with a blackbody emission together with a Comptonized emission. The \nus\ spectra detected a broad Fe emission feature with broadness of $\sim 0.5$ keV centered at energy $\sim 6.5$ keV. 
Additionally, a large excess in tail above 20 keV was observed, compatible with the Compton hump. 

The broad Fe K$\alpha$ line with Compton hump implies the reflection of hard X-ray photons in the accretion disc where the strong velocity field smears discrete features and relativistic effects distort their shapes \citep{Fabian1989}. The self-consistent relativistic reflection model \texttt{relxillCP} successfully modeled the iron line as well as the reflection hump features. From the best fit, a disc inclination of $32.3^{\circ +4.8}_{-4.7}$ with respect to our line of sight is estimated. 
A similar value of disc inclination ($27^{\circ}-34^{\circ}$) was observed for intermittent AMXP HETE J1900.1--2455 \citep{Papitto2013}.

From the reflection model, we have found the iron abundance $A_{Fe}= 2.6^{+2.3}_{-1.6}$ relative to solar values. Although the uncertainty obtained on $A_{Fe}$ is large, the lower limits are consistent with solar abundance. Overabundance of Fe have been reported in some NS and black hole X-ray binaries with \texttt{relxill} model \citep{Degenaar2017, Ludlam2017b, Ludlam2018, Garcia2018}. 

Generally, the AMXPs spectrum are found to be hard with electron temperature of 30--50 keV \citep[e.g.,][]{Falanga2005, Gierlinski2005, Wilkinson2011, Papitto2010, Papitto2013, Sanna2018a, Sanna2018b}. \sax\ was observed in the hard state during outburst of 1998 and 2017 \citep{intZand1999, Pintore2018}.
Our study suggests that the source was observed in the soft state during \nus\ observation similar to the \xm\ observation \citep{Pintore} carried out six days later. The electron temperature of Comptonized plasma ($\sim 2.5$ keV) is somewhat higher than \xm\ observation ($\sim 2.0$ keV).
Also, the energy of Fe emission line ($\sim 6.5$ keV) is lower than the value $\sim 6.8$ keV, which means that the disc was less ionized during \nus\ observation. 
Our estimated value of ionization parameter $\xi \sim 1100$ erg cm s$^{-1}$ is lower than the estimated value, $\xi \sim 1600$ erg cm s$^{-1}$, of \citet{Pintore}.

The optical depth of the Comptonizing medium, $\tau$, can be estimated from the power-law photon index, $\Gamma$, and the electron temperature, $kT_e$, by using the relation
\begin{equation*}
\Gamma = -{1 \over 2} + \sqrt{ {9 \over 4} + { 1 \over {kT_e \over m_ec^2} (1 + {\tau \over 3}) \tau }}
\end{equation*}
\citep[see,][]{Zdziarski}. Using the best fit value of $\Gamma$ and $kT_e$, we have estimated $\tau \simeq 12.6$.
We have estimated the Compton parameter ($y-$parameter) defined as $y = 4 kT_e \tau^2/m_e c^2$ (relative energy gained by the photons in the inverse Compton scattering). Using best fit values, the $y-$parameter is found to be $\sim 3.1$. 

We also estimated the average electron number density ($n_e$) of the Comptonizing region using relation $\tau = \sigma_T n_e l$, where $\sigma_T$ is the Thomson cross-section and $l$ is the geometrical size of the Comptonization cloud. Assuming that the Comptonizing region extends from the NS surface upto the observed inner radius of accretion disc, we found the lower limit of $n_e >3.3 \times 10^{18}$ cm$^{-3}$. On the other hand, the electron number density associated with the reflecting skin above the accretion disc can be estimated using relation $\xi = L/(n_e r^2)$, where $L$ is the unabsorbed incident luminosity, $\xi$ is ionization parameter and $r$ is the inner radius of disc where the reflection component originates. We estimated the unabsorbed incident luminosity using $refl_{frac}/(1+refl_{frac})$ times the unabsorbed luminosity of \texttt{relxillCP} component. Using $log ~\xi = 3.05$, $L \sim 6.7 \times 10^{36}$ erg s$^{-1}$ and $r<57$ km, we obtained $n_e > 1.9 \times 10^{20}$ cm$^{-3}$. 

Our spectral fit can only estimate the upper limit on inner accretion disc radius $R_{in} <5.2 R_{ISCO}$ at 90\% confidence level. For a spinning NS with $a = 0.208$, $R_{ISCO}$ can be approximated using $R_{ISCO} \simeq 6 R_g (1 - 0.54 a)$ \citep{Miller1998}. This implies 
$R_{in} \leq 27.7 ~R_g \leq 57$ km for a mass of 1.4 $M_{\sun}$. Our estimate of the inner disc radius is compatible with the value of 20--40 $R_g$ inferred by \citet{Pintore}. At $1 \sigma$ significance level, we found the inner disc radius of $8-17.6 ~R_g ~(16.6-36.5$ km). This suggests that the disc is probably truncated moderately away from the NS surface. 
Also, given the presence of X-ray pulsation during the outburst of 2015 \citep{Sanna}, the co-rotation radius can be estimated using $R_{co}= (GM/4 \pi^2 \nu^2)^{1/3}$ \citep{Degenaar2017}. Considering that the accretion disc can not be truncated outside the co-rotation radius, $R_{in}$ should be less than 29 km (14 $R_g$) for NS mass of 1.4 $M_{\sun}$, which is consistent with our estimated limit.
The truncated inner disc has been inferred in AMXPs with $R_{in} \simeq 6-15 ~R_g$ \citep{Papitto2009, Cackett2010, Ludlam2017c} and also with larger radii $\sim 15-40 ~R_g$ \citep{Papitto2010, Papitto2013, Pintore}. 

It has been observed that in some LMXBs, the disc truncation occurs at moderate radii due to the pressure exerted by the magnetic field of the NS \citep{Cackett2009, Degenaar2014}. If it is truncated at magnetospheric radius, we can estimate the magnetic field strength. We have used the following expression for the calculation of the magnetic dipole moment \citep{Ibragimov2009}, 

\begin{equation*}
\mu_{25} = 1.36 k_A^{-7/4} \Big( \frac{M}{1.4 M_{\sun}} \Big)^{1/4} \Big(\frac{R_{in}}{10 km} \Big)^{7/4} 
\end{equation*}
\begin{equation*}
\hspace{0.5cm} \times \big(\frac{f_{ang}}{\eta} \frac{F}{10^{-9} ~erg~cm^{-2}~ s^{-1}}\big)^{1/2} \frac{D}{8.5 kpc}
\end{equation*}
where $\mu_{25} = \mu/10^{25}$ G cm$^3$, $\eta$ is the accretion efficiency in the Schwarzchild metric, $f_{ang}$ is the anisotropy correction \citep[which is close to unity;][]{Ibragimov2009} and $k_A$ is a geometry coefficient expected to be $\simeq 0.5-1.1$ \citep{Psaltis1999,Long2005,Kluzniak2007}. 
We assumed $f_{ang} = 1, k_A = 1$ and $\eta = 0.1$ \citep{Cackett2009,Degenaar2017}. 
We then obtained $\mu < 1.7 \times 10^{27}$ G cm$^3$ for $R_{in} <$ 57 km, this leads to a magnetic field strength of $B < 3.4 \times 10^9$ G at poles for NS radius of 10 km ($B \sim 0.4-1.6 \times 10^9$ G at $1\sigma$ significance level). Our estimate of magnetic field strength is within the range determined by \citet{Mukherjee2015}.

\section*{Acknowledgements}
The authors are thankful to Thomas Duaser for discussion on \texttt{relxill} model.
This research has made use of the NuSTAR Data Analysis Software (\textsc{nustardas}) jointly developed by the ASI Science Data Center (ASDC, Italy) and the California Institute of Technology (USA). 
The data was obtained from the High Energy Astrophysics Science Archive Research Center (HEASARC), provided by NASA's Goddard Space Flight Center. RS acknowledges the financial support from the University Grants Commission (UGC), India, under the SRF scheme. 

%%%%%%%%%%%%%%%%%%%%%%%%%%%%%%%%%%%%%%%%%%%%%%%%%%



\bibliographystyle{mnras}
%\bibliography{sample} % if your bibtex file is called example.bib

\begin{thebibliography}{}
\makeatletter
\relax
\def\mn@urlcharsother{\let\do\@makeother \do\$\do\&\do\#\do\^\do\_\do\%\do\~}
\def\mn@doi{\begingroup\mn@urlcharsother \@ifnextchar [ {\mn@doi@}
  {\mn@doi@[]}}
\def\mn@doi@[#1]#2{\def\@tempa{#1}\ifx\@tempa\@empty \href
  {http://dx.doi.org/#2} {doi:#2}\else \href {http://dx.doi.org/#2} {#1}\fi
  \endgroup}
\def\mn@eprint#1#2{\mn@eprint@#1:#2::\@nil}
\def\mn@eprint@arXiv#1{\href {http://arxiv.org/abs/#1} {{\tt arXiv:#1}}}
\def\mn@eprint@dblp#1{\href {http://dblp.uni-trier.de/rec/bibtex/#1.xml}
  {dblp:#1}}
\def\mn@eprint@#1:#2:#3:#4\@nil{\def\@tempa {#1}\def\@tempb {#2}\def\@tempc
  {#3}\ifx \@tempc \@empty \let \@tempc \@tempb \let \@tempb \@tempa \fi \ifx
  \@tempb \@empty \def\@tempb {arXiv}\fi \@ifundefined
  {mn@eprint@\@tempb}{\@tempb:\@tempc}{\expandafter \expandafter \csname
  mn@eprint@\@tempb\endcsname \expandafter{\@tempc}}}

\bibitem[\protect\citeauthoryear{{Altamirano}, {Casella}, {Patruno}, {Wijnands}
   \& {van der Klis}}{{Altamirano} et~al.}{2008}]{Altamirano2008}
{Altamirano} D.,  {Casella} P.,  {Patruno} A.,  {Wijnands} R.,   {van der Klis}
  M.,  2008, \mn@doi [\apjl] {10.1086/528983}, \href
  {http://adsabs.harvard.edu/abs/2008ApJ...674L..45A} {674, L45}

\bibitem[\protect\citeauthoryear{{Arnaud}}{{Arnaud}}{1996}]{Arnaud}
{Arnaud} K.~A.,  1996, in {Jacoby} G.~H.,  {Barnes} J.,  eds,  Astronomical
  Society of the Pacific Conference Series Vol. 101, Astronomical Data Analysis
  Software and Systems V. p.~17

\bibitem[\protect\citeauthoryear{{Braje}, {Romani}  \& {Rauch}}{{Braje}
  et~al.}{2000}]{Braje2000}
{Braje} T.~M.,  {Romani} R.~W.,   {Rauch} K.~P.,  2000, \mn@doi [\apj]
  {10.1086/308448}, \href {http://adsabs.harvard.edu/abs/2000ApJ...531..447B}
  {531, 447}

\bibitem[\protect\citeauthoryear{{Cackett}, {Altamirano}, {Patruno}, {Miller},
  {Reynolds}, {Linares}  \& {Wijnands}}{{Cackett} et~al.}{2009}]{Cackett2009}
{Cackett} E.~M.,  {Altamirano} D.,  {Patruno} A.,  {Miller} J.~M.,  {Reynolds}
  M.,  {Linares} M.,   {Wijnands} R.,  2009, \mn@doi [\apjl]
  {10.1088/0004-637X/694/1/L21}, \href
  {http://adsabs.harvard.edu/abs/2009ApJ...694L..21C} {694, L21}

\bibitem[\protect\citeauthoryear{{Cackett} et~al.,}{{Cackett}
  et~al.}{2010}]{Cackett2010}
{Cackett} E.~M.,  et~al., 2010, \mn@doi [\apj] {10.1088/0004-637X/720/1/205},
  \href {http://adsabs.harvard.edu/abs/2010ApJ...720..205C} {720, 205}

\bibitem[\protect\citeauthoryear{{Cadelano}, {Pallanca}, {Ferraro},
  {Dalessandro}, {Lanzoni}  \& {Patruno}}{{Cadelano} et~al.}{2017}]{Cadelano}
{Cadelano} M.,  {Pallanca} C.,  {Ferraro} F.~R.,  {Dalessandro} E.,  {Lanzoni}
  B.,   {Patruno} A.,  2017, \mn@doi [\apj] {10.3847/1538-4357/aa7b7f}, \href
  {http://adsabs.harvard.edu/abs/2017ApJ...844...53C} {844, 53}

\bibitem[\protect\citeauthoryear{{Dauser}, {Wilms}, {Reynolds}  \&
  {Brenneman}}{{Dauser} et~al.}{2010}]{Dauser2010}
{Dauser} T.,  {Wilms} J.,  {Reynolds} C.~S.,   {Brenneman} L.~W.,  2010,
  \mn@doi [\mnras] {10.1111/j.1365-2966.2010.17393.x}, \href
  {https://ui.adsabs.harvard.edu/#abs/2010MNRAS.409.1534D} {409, 1534}

\bibitem[\protect\citeauthoryear{{Dauser}, {Garcia}, {Wilms}, {B{\"o}ck},
  {Brenneman}, {Falanga}, {Fukumura}  \& {Reynolds}}{{Dauser}
  et~al.}{2013}]{Dauser2013}
{Dauser} T.,  {Garcia} J.,  {Wilms} J.,  {B{\"o}ck} M.,  {Brenneman} L.~W.,
  {Falanga} M.,  {Fukumura} K.,   {Reynolds} C.~S.,  2013, \mn@doi [\mnras]
  {10.1093/mnras/sts710}, \href
  {http://adsabs.harvard.edu/abs/2013MNRAS.430.1694D} {430, 1694}

\bibitem[\protect\citeauthoryear{{Dauser}, {Garcia}, {Parker}, {Fabian}  \&
  {Wilms}}{{Dauser} et~al.}{2014}]{Dauser2014}
{Dauser} T.,  {Garcia} J.,  {Parker} M.~L.,  {Fabian} A.~C.,   {Wilms} J.,
  2014, \mn@doi [\mnras] {10.1093/mnrasl/slu125}, \href
  {https://ui.adsabs.harvard.edu/#abs/2014MNRAS.444L.100D} {444, L100}

\bibitem[\protect\citeauthoryear{{Dauser}, {Garc{\'{\i}}a}, {Walton},
  {Eikmann}, {Kallman}, {McClintock}  \& {Wilms}}{{Dauser}
  et~al.}{2016}]{Dauser2016}
{Dauser} T.,  {Garc{\'{\i}}a} J.,  {Walton} D.~J.,  {Eikmann} W.,  {Kallman}
  T.,  {McClintock} J.,   {Wilms} J.,  2016, \mn@doi [\aap]
  {10.1051/0004-6361/201628135}, \href
  {http://adsabs.harvard.edu/abs/2016A%26A...590A..76D} {590, A76}

\bibitem[\protect\citeauthoryear{{Degenaar}, {Miller}, {Harrison}, {Kennea},
  {Kouveliotou}  \& {Younes}}{{Degenaar} et~al.}{2014}]{Degenaar2014}
{Degenaar} N.,  {Miller} J.~M.,  {Harrison} F.~A.,  {Kennea} J.~A.,
  {Kouveliotou} C.,   {Younes} G.,  2014, \mn@doi [\apjl]
  {10.1088/2041-8205/796/1/L9}, \href
  {http://adsabs.harvard.edu/abs/2014ApJ...796L...9D} {796, L9}

\bibitem[\protect\citeauthoryear{{Degenaar}, {Pinto}, {Miller}, {Wijnands},
  {Altamirano}, {Paerels}, {Fabian}  \& {Chakrabarty}}{{Degenaar}
  et~al.}{2017}]{Degenaar2017}
{Degenaar} N.,  {Pinto} C.,  {Miller} J.~M.,  {Wijnands} R.,  {Altamirano} D.,
  {Paerels} F.,  {Fabian} A.~C.,   {Chakrabarty} D.,  2017, \mn@doi [\mnras]
  {10.1093/mnras/stw2355}, \href
  {http://adsabs.harvard.edu/abs/2017MNRAS.464..398D} {464, 398}

\bibitem[\protect\citeauthoryear{{Fabian} \& {Ross}}{{Fabian} \&
  {Ross}}{2010}]{Fabian2010}
{Fabian} A.~C.,  {Ross} R.~R.,  2010, \mn@doi [\ssr]
  {10.1007/s11214-010-9699-y}, \href
  {http://adsabs.harvard.edu/abs/2010SSRv..157..167F} {157, 167}

\bibitem[\protect\citeauthoryear{{Fabian}, {Rees}, {Stella}  \&
  {White}}{{Fabian} et~al.}{1989}]{Fabian1989}
{Fabian} A.~C.,  {Rees} M.~J.,  {Stella} L.,   {White} N.~E.,  1989, \mn@doi
  [\mnras] {10.1093/mnras/238.3.729}, \href
  {http://adsabs.harvard.edu/abs/1989MNRAS.238..729F} {238, 729}

\bibitem[\protect\citeauthoryear{{Falanga} et~al.,}{{Falanga}
  et~al.}{2005}]{Falanga2005}
{Falanga} M.,  et~al., 2005, \mn@doi [\aap] {10.1051/0004-6361:20053472}, \href
  {http://adsabs.harvard.edu/abs/2005A%26A...444...15F} {444, 15}

\bibitem[\protect\citeauthoryear{{Garc{\'\i}a} \& {Kallman}}{{Garc{\'\i}a} \&
  {Kallman}}{2010}]{Garcia2010}
{Garc{\'\i}a} J.,  {Kallman} T.~R.,  2010, \mn@doi [\apj]
  {10.1088/0004-637X/718/2/695}, \href
  {https://ui.adsabs.harvard.edu/#abs/2010ApJ...718..695G} {718, 695}

\bibitem[\protect\citeauthoryear{{Garc{\'\i}a}, {Dauser}, {Reynolds},
  {Kallman}, {McClintock}, {Wilms}  \& {Eikmann}}{{Garc{\'\i}a}
  et~al.}{2013}]{Garcia2013}
{Garc{\'\i}a} J.,  {Dauser} T.,  {Reynolds} C.~S.,  {Kallman} T.~R.,
  {McClintock} J.~E.,  {Wilms} J.,   {Eikmann} W.,  2013, \mn@doi [\apj]
  {10.1088/0004-637X/768/2/146}, \href
  {https://ui.adsabs.harvard.edu/#abs/2013ApJ...768..146G} {768, 146}

\bibitem[\protect\citeauthoryear{{Garc{\'\i}a} et~al.,}{{Garc{\'\i}a}
  et~al.}{2014}]{Garcia2014}
{Garc{\'\i}a} J.,  et~al., 2014, \mn@doi [\apj] {10.1088/0004-637X/782/2/76},
  \href {https://ui.adsabs.harvard.edu/#abs/2014ApJ...782...76G} {782, 76}

\bibitem[\protect\citeauthoryear{{Garc{\'{\i}}a} et~al.,}{{Garc{\'{\i}}a}
  et~al.}{2018}]{Garcia2018}
{Garc{\'{\i}}a} J.~A.,  et~al., 2018, preprint, \href
  {http://adsabs.harvard.edu/abs/2018arXiv180701949G} {} (\mn@eprint {arXiv}
  {1807.01949})

\bibitem[\protect\citeauthoryear{{Ghosh}}{{Ghosh}}{2007}]{Ghosh}
{Ghosh} P.,  2007, {Rotation and Accretion Powered Pulsars}.
World Scientific Publishing Co, \mn@doi{10.1142/4806}

\bibitem[\protect\citeauthoryear{{Gierli{\'n}ski} \&
  {Poutanen}}{{Gierli{\'n}ski} \& {Poutanen}}{2005}]{Gierlinski2005}
{Gierli{\'n}ski} M.,  {Poutanen} J.,  2005, \mn@doi [\mnras]
  {10.1111/j.1365-2966.2005.09004.x}, \href
  {http://adsabs.harvard.edu/abs/2005MNRAS.359.1261G} {359, 1261}

\bibitem[\protect\citeauthoryear{{Harrison} et~al.,}{{Harrison}
  et~al.}{2013}]{Harrison}
{Harrison} F.~A.,  et~al., 2013, \mn@doi [\apj] {10.1088/0004-637X/770/2/103},
  \href {http://adsabs.harvard.edu/abs/2013ApJ...770..103H} {770, 103}

\bibitem[\protect\citeauthoryear{{Ibragimov} \& {Poutanen}}{{Ibragimov} \&
  {Poutanen}}{2009}]{Ibragimov2009}
{Ibragimov} A.,  {Poutanen} J.,  2009, \mn@doi [\mnras]
  {10.1111/j.1365-2966.2009.15477.x}, \href
  {http://adsabs.harvard.edu/abs/2009MNRAS.400..492I} {400, 492}

\bibitem[\protect\citeauthoryear{{in 't Zand} et~al.,}{{in 't Zand}
  et~al.}{1999}]{intZand1999}
{in 't Zand} J.~J.~M.,  et~al., 1999, \aap, \href
  {http://adsabs.harvard.edu/abs/1999A%26A...345..100I} {345, 100}

\bibitem[\protect\citeauthoryear{{in 't Zand}, {van Kerkwijk}, {Pooley},
  {Verbunt}, {Wijnands}  \& {Lewin}}{{in 't Zand} et~al.}{2001}]{intZand2001}
{in 't Zand} J.~J.~M.,  {van Kerkwijk} M.~H.,  {Pooley} D.,  {Verbunt} F.,
  {Wijnands} R.,   {Lewin} W.~H.~G.,  2001, \mn@doi [\apjl] {10.1086/338361},
  \href {http://adsabs.harvard.edu/abs/2001ApJ...563L..41I} {563, L41}

\bibitem[\protect\citeauthoryear{{Klu{\'z}niak} \& {Rappaport}}{{Klu{\'z}niak}
  \& {Rappaport}}{2007}]{Kluzniak2007}
{Klu{\'z}niak} W.,  {Rappaport} S.,  2007, \mn@doi [\apj] {10.1086/522954},
  \href {http://adsabs.harvard.edu/abs/2007ApJ...671.1990K} {671, 1990}

\bibitem[\protect\citeauthoryear{{Kolehmainen}, {Done}  \& {D{\'{\i}}az
  Trigo}}{{Kolehmainen} et~al.}{2011}]{Kolehmainen2011}
{Kolehmainen} M.,  {Done} C.,   {D{\'{\i}}az Trigo} M.,  2011, \mn@doi [\mnras]
  {10.1111/j.1365-2966.2011.19040.x}, \href
  {http://adsabs.harvard.edu/abs/2011MNRAS.416..311K} {416, 311}

\bibitem[\protect\citeauthoryear{{Kuulkers}, {den Hartog}, {in't Zand},
  {Verbunt}, {Harris}  \& {Cocchi}}{{Kuulkers} et~al.}{2003}]{Kuulkers2003}
{Kuulkers} E.,  {den Hartog} P.~R.,  {in't Zand} J.~J.~M.,  {Verbunt} F.~W.~M.,
   {Harris} W.~E.,   {Cocchi} M.,  2003, \mn@doi [\aap]
  {10.1051/0004-6361:20021781}, \href
  {http://adsabs.harvard.edu/abs/2003A%26A...399..663K} {399, 663}

\bibitem[\protect\citeauthoryear{{Kuulkers} et~al.,}{{Kuulkers}
  et~al.}{2015}]{Kuulkers-detect}
{Kuulkers} E.,  et~al., 2015, The Astronomer's Telegram, \href
  {http://adsabs.harvard.edu/abs/2015ATel.7098....1K} {7098}

\bibitem[\protect\citeauthoryear{{Long}, {Romanova}  \& {Lovelace}}{{Long}
  et~al.}{2005}]{Long2005}
{Long} M.,  {Romanova} M.~M.,   {Lovelace} R.~V.~E.,  2005, \mn@doi [\apj]
  {10.1086/497000}, \href {http://adsabs.harvard.edu/abs/2005ApJ...634.1214L}
  {634, 1214}

\bibitem[\protect\citeauthoryear{{Ludlam} et~al.,}{{Ludlam}
  et~al.}{2017a}]{Ludlam2017b}
{Ludlam} R.~M.,  et~al., 2017a, \mn@doi [\apj] {10.3847/1538-4357/836/1/140},
  \href {http://adsabs.harvard.edu/abs/2017ApJ...836..140L} {836, 140}

\bibitem[\protect\citeauthoryear{{Ludlam}, {Miller}, {Cackett}, {Degenaar}  \&
  {Bostrom}}{{Ludlam} et~al.}{2017b}]{Ludlam2017a}
{Ludlam} R.~M.,  {Miller} J.~M.,  {Cackett} E.~M.,  {Degenaar} N.,   {Bostrom}
  A.~C.,  2017b, \mn@doi [\apj] {10.3847/1538-4357/aa661a}, \href
  {http://adsabs.harvard.edu/abs/2017ApJ...838...79L} {838, 79}

\bibitem[\protect\citeauthoryear{{Ludlam}, {Miller}, {Degenaar}, {Sanna},
  {Cackett}, {Altamirano}  \& {King}}{{Ludlam} et~al.}{2017c}]{Ludlam2017c}
{Ludlam} R.~M.,  {Miller} J.~M.,  {Degenaar} N.,  {Sanna} A.,  {Cackett} E.~M.,
   {Altamirano} D.,   {King} A.~L.,  2017c, \mn@doi [\apj]
  {10.3847/1538-4357/aa8b1b}, \href
  {http://adsabs.harvard.edu/abs/2017ApJ...847..135L} {847, 135}

\bibitem[\protect\citeauthoryear{{Ludlam} et~al.,}{{Ludlam}
  et~al.}{2018}]{Ludlam2018}
{Ludlam} R.~M.,  et~al., 2018, \mn@doi [\apjl] {10.3847/2041-8213/aabee6},
  \href {http://adsabs.harvard.edu/abs/2018ApJ...858L...5L} {858, L5}

\bibitem[\protect\citeauthoryear{{Markwardt} \& {Swank}}{{Markwardt} \&
  {Swank}}{2005}]{Markwardt2005}
{Markwardt} C.~B.,  {Swank} J.~H.,  2005, The Astronomer's Telegram, \href
  {http://adsabs.harvard.edu/abs/2005ATel..495....1M} {495}

\bibitem[\protect\citeauthoryear{{Miller}, {Lamb}  \& {Cook}}{{Miller}
  et~al.}{1998}]{Miller1998}
{Miller} M.~C.,  {Lamb} F.~K.,   {Cook} G.~B.,  1998, \mn@doi [\apj]
  {10.1086/306533}, \href {http://adsabs.harvard.edu/abs/1998ApJ...509..793M}
  {509, 793}

\bibitem[\protect\citeauthoryear{{Mitsuda} et~al.,}{{Mitsuda}
  et~al.}{1984}]{Mitsuda}
{Mitsuda} K.,  et~al., 1984, \pasj, \href
  {http://adsabs.harvard.edu/abs/1984PASJ...36..741M} {36, 741}

\bibitem[\protect\citeauthoryear{{Mondal}, {Pahari}, {Dewangan}, {Misra}  \&
  {Raychaudhuri}}{{Mondal} et~al.}{2017}]{Mondal2017}
{Mondal} A.~S.,  {Pahari} M.,  {Dewangan} G.~C.,  {Misra} R.,   {Raychaudhuri}
  B.,  2017, \mn@doi [\mnras] {10.1093/mnras/stx039}, \href
  {http://adsabs.harvard.edu/abs/2017MNRAS.466.4991M} {466, 4991}

\bibitem[\protect\citeauthoryear{{Mukherjee}, {Bult}, {van der Klis}  \&
  {Bhattacharya}}{{Mukherjee} et~al.}{2015}]{Mukherjee2015}
{Mukherjee} D.,  {Bult} P.,  {van der Klis} M.,   {Bhattacharya} D.,  2015,
  \mn@doi [\mnras] {10.1093/mnras/stv1542}, \href
  {http://adsabs.harvard.edu/abs/2015MNRAS.452.3994M} {452, 3994}

\bibitem[\protect\citeauthoryear{{Negoro} et~al.,}{{Negoro}
  et~al.}{2017}]{Negoro2017}
{Negoro} H.,  et~al., 2017, The Astronomer's Telegram, \href
  {http://adsabs.harvard.edu/abs/2017ATel10821....1N} {10821}

\bibitem[\protect\citeauthoryear{{Ortolani}, {Barbuy}  \& {Bica}}{{Ortolani}
  et~al.}{1994}]{Ortolani1994}
{Ortolani} S.,  {Barbuy} B.,   {Bica} E.,  1994, \aaps, \href
  {http://adsabs.harvard.edu/abs/1994A%26AS..108..653O} {108, 653}

\bibitem[\protect\citeauthoryear{{Papitto}, {Di Salvo}, {D'A{\`i}}, {Iaria},
  {Burderi}, {Riggio}, {Menna}  \& {Robba}}{{Papitto}
  et~al.}{2009}]{Papitto2009}
{Papitto} A.,  {Di Salvo} T.,  {D'A{\`i}} A.,  {Iaria} R.,  {Burderi} L.,
  {Riggio} A.,  {Menna} M.~T.,   {Robba} N.~R.,  2009, \mn@doi [\aap]
  {10.1051/0004-6361:200811401}, \href
  {http://adsabs.harvard.edu/abs/2009A%26A...493L..39P} {493, L39}

\bibitem[\protect\citeauthoryear{{Papitto}, {Riggio}, {di Salvo}, {Burderi},
  {D'A{\`i}}, {Iaria}, {Bozzo}  \& {Menna}}{{Papitto}
  et~al.}{2010}]{Papitto2010}
{Papitto} A.,  {Riggio} A.,  {di Salvo} T.,  {Burderi} L.,  {D'A{\`i}} A.,
  {Iaria} R.,  {Bozzo} E.,   {Menna} M.~T.,  2010, \mn@doi [\mnras]
  {10.1111/j.1365-2966.2010.17090.x}, \href
  {http://adsabs.harvard.edu/abs/2010MNRAS.407.2575P} {407, 2575}

\bibitem[\protect\citeauthoryear{{Papitto} et~al.,}{{Papitto}
  et~al.}{2013}]{Papitto2013}
{Papitto} A.,  et~al., 2013, \mn@doi [\mnras] {10.1093/mnras/sts605}, \href
  {http://adsabs.harvard.edu/abs/2013MNRAS.429.3411P} {429, 3411}

\bibitem[\protect\citeauthoryear{{Patruno} \& {Watts}}{{Patruno} \&
  {Watts}}{2012}]{Patruno2012}
{Patruno} A.,  {Watts} A.~L.,  2012, preprint, \href
  {http://adsabs.harvard.edu/abs/2012arXiv1206.2727P} {} (\mn@eprint {arXiv}
  {1206.2727})

\bibitem[\protect\citeauthoryear{{Patruno}, {Altamirano}, {Hessels}, {Casella},
  {Wijnands}  \& {van der Klis}}{{Patruno} et~al.}{2009}]{Patruno2009}
{Patruno} A.,  {Altamirano} D.,  {Hessels} J.~W.~T.,  {Casella} P.,  {Wijnands}
  R.,   {van der Klis} M.,  2009, \mn@doi [\apj]
  {10.1088/0004-637X/690/2/1856}, \href
  {http://adsabs.harvard.edu/abs/2009ApJ...690.1856P} {690, 1856}

\bibitem[\protect\citeauthoryear{{Patruno} et~al.,}{{Patruno}
  et~al.}{2010}]{Patruno2010}
{Patruno} A.,  et~al., 2010, The Astronomer's Telegram, \href
  {http://adsabs.harvard.edu/abs/2010ATel.2407....1P} {2407}

\bibitem[\protect\citeauthoryear{{Pintore} et~al.,}{{Pintore}
  et~al.}{2016}]{Pintore}
{Pintore} F.,  et~al., 2016, \mn@doi [\mnras] {10.1093/mnras/stw176}, \href
  {http://adsabs.harvard.edu/abs/2016MNRAS.457.2988P} {457, 2988}

\bibitem[\protect\citeauthoryear{{Pintore} et~al.,}{{Pintore}
  et~al.}{2018}]{Pintore2018}
{Pintore} F.,  et~al., 2018, \mn@doi [\mnras] {10.1093/mnras/sty1735}, \href
  {http://adsabs.harvard.edu/abs/2018MNRAS.479.4084P} {479, 4084}

\bibitem[\protect\citeauthoryear{{Poutanen}}{{Poutanen}}{2006}]{Poutanen2006}
{Poutanen} J.,  2006, \mn@doi [Advances in Space Research]
  {10.1016/j.asr.2006.04.025}, \href
  {http://adsabs.harvard.edu/abs/2006AdSpR..38.2697P} {38, 2697}

\bibitem[\protect\citeauthoryear{{Psaltis} \& {Chakrabarty}}{{Psaltis} \&
  {Chakrabarty}}{1999}]{Psaltis1999}
{Psaltis} D.,  {Chakrabarty} D.,  1999, \mn@doi [\apj] {10.1086/307525}, \href
  {http://adsabs.harvard.edu/abs/1999ApJ...521..332P} {521, 332}

\bibitem[\protect\citeauthoryear{{Ross} \& {Fabian}}{{Ross} \&
  {Fabian}}{2005}]{Ross2005}
{Ross} R.~R.,  {Fabian} A.~C.,  2005, \mn@doi [\mnras]
  {10.1111/j.1365-2966.2005.08797.x}, \href
  {http://adsabs.harvard.edu/abs/2005MNRAS.358..211R} {358, 211}

\bibitem[\protect\citeauthoryear{{Ross}, {Fabian}  \& {Young}}{{Ross}
  et~al.}{1999}]{Ross1999}
{Ross} R.~R.,  {Fabian} A.~C.,   {Young} A.~J.,  1999, \mn@doi [\mnras]
  {10.1046/j.1365-8711.1999.02528.x}, \href
  {http://adsabs.harvard.edu/abs/1999MNRAS.306..461R} {306, 461}

\bibitem[\protect\citeauthoryear{{Sanna} et~al.,}{{Sanna} et~al.}{2016}]{Sanna}
{Sanna} A.,  et~al., 2016, \mn@doi [\mnras] {10.1093/mnras/stw740}, \href
  {http://adsabs.harvard.edu/abs/2016MNRAS.459.1340S} {459, 1340}

\bibitem[\protect\citeauthoryear{{Sanna} et~al.,}{{Sanna}
  et~al.}{2018a}]{Sanna2018a}
{Sanna} A.,  et~al., 2018a, preprint, \href
  {http://adsabs.harvard.edu/abs/2018arXiv180806796S} {} (\mn@eprint {arXiv}
  {1808.06796})

\bibitem[\protect\citeauthoryear{{Sanna} et~al.,}{{Sanna}
  et~al.}{2018b}]{Sanna2018b}
{Sanna} A.,  et~al., 2018b, \mn@doi [\aap] {10.1051/0004-6361/201732262}, \href
  {http://adsabs.harvard.edu/abs/2018A%26A...610L...2S} {610, L2}

\bibitem[\protect\citeauthoryear{{Verner}, {Ferland}, {Korista}  \&
  {Yakovlev}}{{Verner} et~al.}{1996}]{Verner}
{Verner} D.~A.,  {Ferland} G.~J.,  {Korista} K.~T.,   {Yakovlev} D.~G.,  1996,
  \mn@doi [\apj] {10.1086/177435}, \href
  {http://adsabs.harvard.edu/abs/1996ApJ...465..487V} {465, 487}

\bibitem[\protect\citeauthoryear{{Wang}, {M{\'e}ndez}, {Sanna}, {Altamirano}
  \& {Belloni}}{{Wang} et~al.}{2017}]{Wang2017}
{Wang} Y.,  {M{\'e}ndez} M.,  {Sanna} A.,  {Altamirano} D.,   {Belloni} T.~M.,
  2017, \mn@doi [\mnras] {10.1093/mnras/stx671}, \href
  {http://adsabs.harvard.edu/abs/2017MNRAS.468.2256W} {468, 2256}

\bibitem[\protect\citeauthoryear{{Wilkins} \& {Fabian}}{{Wilkins} \&
  {Fabian}}{2012}]{Wilkins2012}
{Wilkins} D.~R.,  {Fabian} A.~C.,  2012, \mn@doi [\mnras]
  {10.1111/j.1365-2966.2012.21308.x}, \href
  {http://adsabs.harvard.edu/abs/2012MNRAS.424.1284W} {424, 1284}

\bibitem[\protect\citeauthoryear{{Wilkinson}, {Patruno}, {Watts}  \&
  {Uttley}}{{Wilkinson} et~al.}{2011}]{Wilkinson2011}
{Wilkinson} T.,  {Patruno} A.,  {Watts} A.,   {Uttley} P.,  2011, \mn@doi
  [\mnras] {10.1111/j.1365-2966.2010.17532.x}, \href
  {http://adsabs.harvard.edu/abs/2011MNRAS.410.1513W} {410, 1513}

\bibitem[\protect\citeauthoryear{{Wilms}, {Allen}  \& {McCray}}{{Wilms}
  et~al.}{2000}]{Wilms}
{Wilms} J.,  {Allen} A.,   {McCray} R.,  2000, \mn@doi [\apj] {10.1086/317016},
  \href {http://adsabs.harvard.edu/abs/2000ApJ...542..914W} {542, 914}

\bibitem[\protect\citeauthoryear{{Zdziarski}, {Johnson}  \&
  {Magdziarz}}{{Zdziarski} et~al.}{1996}]{Zdziarski}
{Zdziarski} A.~A.,  {Johnson} W.~N.,   {Magdziarz} P.,  1996, \mn@doi [\mnras]
  {10.1093/mnras/283.1.193}, \href
  {http://adsabs.harvard.edu/abs/1996MNRAS.283..193Z} {283, 193}

\bibitem[\protect\citeauthoryear{{{\.Z}ycki}, {Done}  \& {Smith}}{{{\.Z}ycki}
  et~al.}{1999}]{Zycki}
{{\.Z}ycki} P.~T.,  {Done} C.,   {Smith} D.~A.,  1999, \mn@doi [\mnras]
  {10.1046/j.1365-8711.1999.02885.x}, \href
  {http://adsabs.harvard.edu/abs/1999MNRAS.309..561Z} {309, 561}

\makeatother
\end{thebibliography}

%\begin{thebibliography}{99}
%\end{thebibliography}

%%%%%%%%%%%%%%%%%%%%%%%%%%%%%%%%%%%%%%%%%%%%%%%%%%

%%%%%%%%%%%%%%%%% APPENDICES %%%%%%%%%%%%%%%%%%%%%

%\appendix


%%%%%%%%%%%%%%%%%%%%%%%%%%%%%%%%%%%%%%%%%%%%%%%%%%

% Don't change these lines
\bsp	% typesetting comment
\label{lastpage}
\end{document}